# Monolithic integration of InP on Si by molten alloy driven selective area epitaxial growth

Dmitrii V. Viazmitinov[a,‡], Yury Berdnikov[b,‡], Shima Kadkhodazadeh[c], Anna Dragunova[d], Nickolay Sibirev[b], Natalia Kryzhanovskaya[d], Ilya Radko[e], Alexander Huck[e], Kresten Yvind[a] and Elizaveta Semenova[a]

- [a.] DTU Fotonik, Technical University of Denmark, Kongens Lyngby DK-2800, Denmark.
- [b.] ITMO University, Kronverkskiy 49, 197101 St. Petersburg, Russia.
- [c.] DTU Nanolab–National Centre for Nano Fabrication and Characterization, Technical University of Denmark, Kongens Lyngby DK-2800, Denmark.
- [d.] National Research University Higher School of Economics, 16 Soyuza Pechatnikov, St Petersburg 190008, Russia.
- [e.] Center for Macroscopic Quantum States (bigQ), Department of Physics, Technical University of Denmark, Kongens Lyngby DK-2800, Denmark.

‡ These authors contributed equally.

## Abstract

We report a new approach for monolithic integration of III-V materials into silicon, based on selective area growth and driven by a molten alloy in metal-organic vapor epitaxy. Our method includes elements of both selective area and droplet-mediated growths and combines the advantages of the two techniques. Using this approach, we obtain organized arrays of high crystalline quality InP insertions into (100) oriented Si substrates. Our detailed structural, morphological and optical studies reveal the conditions leading to defect formation. These conditions are then eliminated to optimize the process for obtaining dislocation-free InP nanostructures grown directly on Si and buried below the top surface. The PL signal from these structures exhibits a narrow peak at the InP bandgap energy. The fundamental aspects of the growth are studied by modeling the InP nucleation process. The model is fitted by our x-ray diffraction measurements and correlates well with the results of our transmission electron microscopy and optical investigations. Our method constitutes a new approach for the monolithic integration of active III-V material into Si platform and opens up new opportunities in active Si photonics.

## Introduction

In virtue of indisputable advantages, such as robustness, sustainable oxide and wide prevalence in nature, Si-based photonics and electronics empowered by complementary-metal-oxide-semiconductor (CMOS) technology are being continuously developed with the aim of lowering power consumption, high-speed operation and continual device miniaturization [1–3]. Since the scaling on silicon devices approaches the physical limit, new materials need to be integrated on to the Si platform, to ensure further developments in electronics and to achieve new functionalities in photonic applications[3]. III-V materials are, for example, required for active photonic and high-speed electronic applications, due to their direct band gap structures and high electron mobility[4]. This makes monolithic integration of III-V materials onto the Si platform very attractive[5], as it will have a considerable impact on both research and the industrial production of novel integrated photonics[6–9], optical interconnects[10] and high-speed electronic devices[11,12]. Such an integration, however, is challenging, due to the large lattice mismatch between Si and III-V materials, their highly dissimilar thermal expansion coefficients and polar-on-nonpolar epitaxy problems, which typically lead to structures with defect density levels above what is acceptable for device fabrication[13].

In recent years tremendous effort has been devoted to implementing techniques that reduce the number of defects in the III-V nanostructures monolithically integrated into Si. They include wafer bonding[14–17], transfer printing[18] and direct epitaxial methods of integration[13,19], such as growth on thick buffer layers[20–22], various selective-area growth (SAG) methods[12,23–25] and different approaches utilizing liquid group III metallic droplet as a growth catalyst [26–28]. The latter two families of techniques are more promising, due to their flexible integration into Si-based circuits and avoiding usage of III-V substrates, complicated wafer bonding techniques and difficulties in alignment of III-V elements to the Si-circuit. They have subsequently already been used to obtain high-quality III-V structures grown on Si for laser applications [23,29–31]. Besides effective dislocation filtering, the SAG approaches have been used to produce highly ordered nanostructure arrays with precise positioning [30,32–34]. The use of group III droplets enables the growth of III-V materials on Si with low defect density, due to a small nucleation area and close to thermodynamic equilibrium growth conditions [26,28,35].

Here, we present a new CMOS compatible epitaxial growth technique by metal-organic vapor phase epitaxy (MOVPE), named molten alloy driven selective-area growth (MADSAG). It is based on using a droplet of the group III element of the final crystal combined with the advantages of SAG.

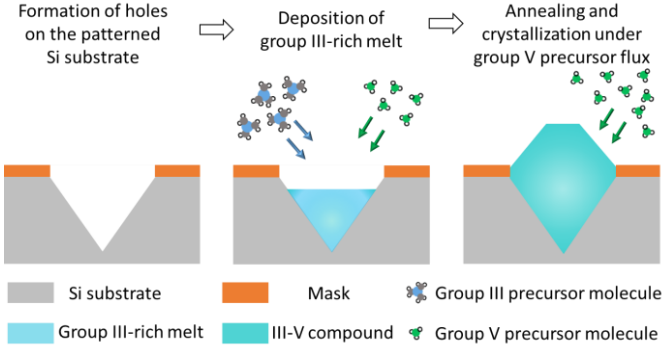

Fig. 1 Schematic for MADSAG experimental procedure

Implementing this method, we obtain high crystalline quality InP nanoinsertions into (100) oriented Si without using a buffer layer. While our MADSAG method can be used to grow different III-V compounds on Si, the choice of materials here is motivated by further applications in telecom-wavelength compatible photonics[36], and the otherwise scarce reports of the epitaxy of InP on Si substrates in the literature.

**Experimental**

MADSAG experimental procedure consists of substrate masking and formation of openings followed by epitaxial deposition of group III-rich melt and annealing under group V precursor flux, as shown in Fig. 1. The samples investigated here were prepared on Si (100) wafers covered by a $Si_3N_4$ mask containing arrays of circular openings 200 nm in diameter and arranged with a pitch of 800 nm. After etching the samples in KOH, inverted pyramidal holes confined by the {111} family of planes are formed in each mask opening. The details of the mask fabrication and Si surface preparation are provided in section S1 of the Supplementary Information (SI). The epitaxial growth on the patterned wafers, in turn, consisted of depositing P-rich In melts with subsequent annealing in a $PH_3$ ambient. Details of the epitaxial growth are described in section S2 of the SI. A series of samples were synthesized at different temperatures and $PH_3$ fluxes, in order to understand the InP nucleation by the MADSAG mechanism, through monitoring the evolution of the morphology, chemical composition and crystal structure and quality of the grown nanoinsertions. The structures were studied in detail using scanning electron microscopy (SEM), high resolution scanning transmission electron microscopy (HR STEM), energy-dispersive X-ray spectroscopy (EDX) and X-ray diffractometry (XRD). Their optical properties were investigated using photoluminescence (PL) measurements.

**Results and discussion**

We implement our approach to obtain organized arrays of high crystalline quality InP insertions into (100) oriented Si substrates. We present and discuss the results of detailed structural, morphological and optical studies of the obtained structures to identify the growth conditions leading to no defect formation. We identify the impact of the $PH_3$ flux and substrate temperature during the annealing to optimize the process for obtaining dislocation-free InP nanostructures grown directly on Si and buried below the top surface.

**Optimization of the group V flux**

We start by comparing three samples grown identically but annealed under different $PH_3$ fluxes, in order to investigate the impact of the $PH_3$ flux during annealing on the nucleation of InP inside the openings in Si(100). While changing the $PH_3$ flux, the substrate temperature was kept at 600 °C. Moreover, the total number of $PH_3$ molecules per unit surface was kept constant during annealing, by adjusting the annealing time. Images of three samples with the deposited P-rich In melts not annealed, and annealed at $1.2 \times 10^{-2}$ mol/min (15 min) and $2.2 \times 10^{-2}$ mol/min (8.1 min) $PH_3$ flux rates are presented in Fig. 2, respectively. Fig. 2a–c show plan-view SEM images of the structures, Fig. 2d – f show STEM images of the cross-sections of the samples, Fig. 2g – i are fast Fourier transforms (FFT) of the HR STEM images of the regions marked in Fig. 2d – f, and Fig. 2j – l are EDX composition maps of the samples. Details of the STEM and EDX measurements along with the HR STEM images corresponding to the FFTs in Fig. 2g – i can be found in section S4 of the SI. We find that InP does not nucleate in the sample without annealing, evident by the FFTs of HR STEM images from this sample matching the body-centered tetragonal (bct) crystal structure of In (Fig. 2g). Since the nucleation of the solid phase is known to be limited by the concentration of group V atoms [37,38], the P concentration in this case could not have been high enough to initiate an irreversible crystallization. The concentration of P of 12% (comparable with the method accuracy), measured by EDX in Fig. 2j, confirms this. Annealing under a $PH_3$ flow is expected to increase the P concentration in the melt, and subsequently promote the nucleation and further crystallization of InP. Images of the sample annealed at $PH_3$ flux $1.2\times10^{-2}$ mol/min suggest a partial nucleation of InP, as regions of both bct In and face centered cubic (fcc) InP can be observed in this sample (see Fig. 2h).

The EDX measurements support this by revealing a non-homogeneous distribution of P and its segregation in the regions with an fcc InP phase (Fig. 2k). The InP region has the same crystallographic orientation as the Si substrate (see section S4 in the SI) and therefore, we conclude that InP nucleates at the interface between the In melt and Si and it inherits the substrate orientation.

Increasing the $PH_3$ flux leads to increase in the volume of the crystalized InP, evident by the images in Fig. 2c, f, i and l, from the sample annealed at $PH_3$ flux $2.2\times10^{-2}$ mol/min. Expansion of the volume of material in the nanoinsertions due to complete conversion of In to InP can be clearly seen in the plan-view SEM image in Fig. 2c. HR STEM images and the corresponding FFTs from this sample only show the fcc InP crystal phase, and the P concentration in the EDX map in Fig. 2l has a uniform distribution. We observe a sharp interface between the InP and Si lattices, with the crystal orientation of InP matched to that of the surrounding Si crystal (see section S4 of SI). Further increasing the $PH_3$ flux beyond this value did not considerably affect the outcome.

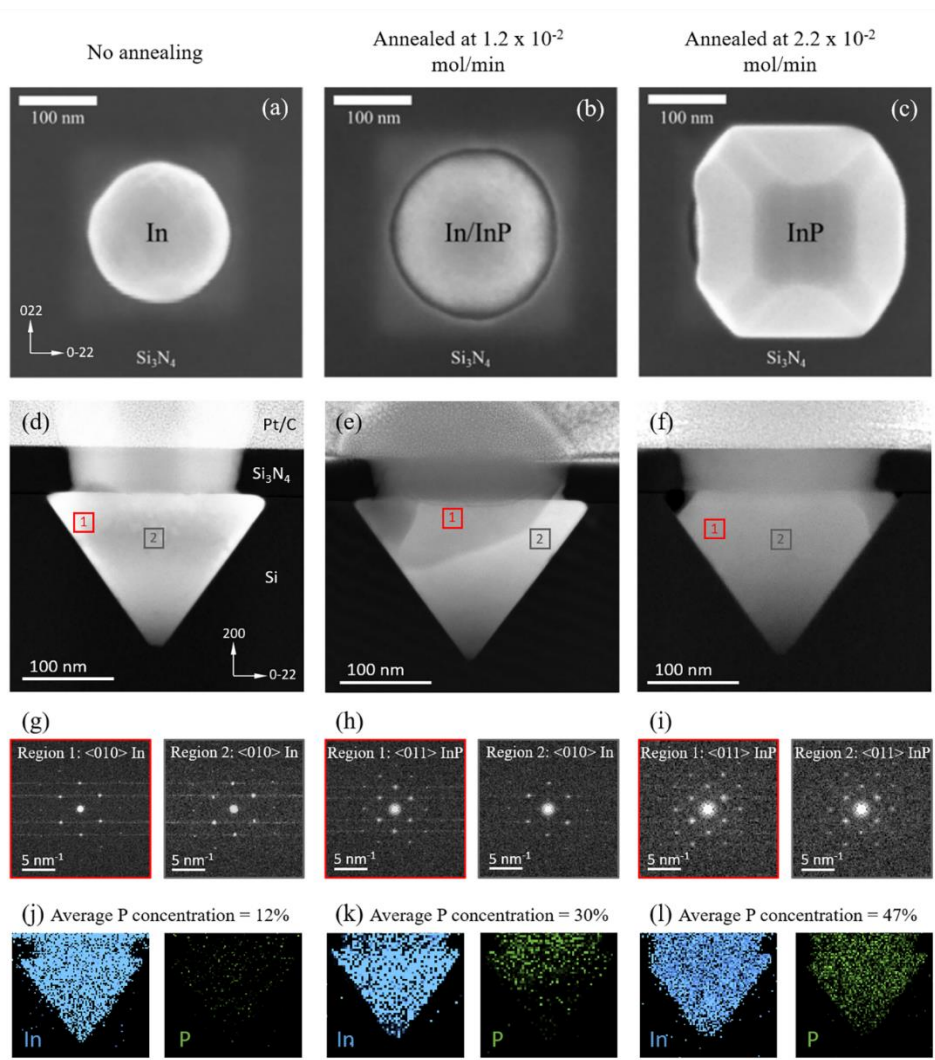

Fig. 2. (a-c) SEM images of nanoinsertion grown in (100) Si and annealed under different conditions in plan-view geometry, (d-f) STEM images of the cross-sections of the nanoinsertions, (g-i) FFTs of HR STEM images of the regions in the nanoinsertions marked in (d-f), and (j-l) the corresponding concentration maps of In and P, illustrating the crystallization during the MADSAG process without annealing (a, d, g, j) and with annealing under PH3 flux of 1.2x10-2 (b, e, h, k) and 2.2x10-2 mol/min (c, f, i, l). The crystallographic directions of the Si substrate in the plan-view and cross-sectional images are shown in (a) and (d). The EDX measurements have an accuracy of 10-15%.

Based on the results we conclude that the epitaxial growth in the MADSAG process consists of two stages: 1. deposition of a group III droplet and 2. nucleation and further crystallization of the III-V material during annealing under a group V flux. It is important to remember that despite the difference in crystallization, the total amount of PH3 molecules per unit surface area during annealing were kept constant. The nucleation and growth of the InP crystal are, however, driven by the concentration of P atoms in the In melt, which is dependent on desorption rates of P in the In melt. Meanwhile, it is well-known that the desorption of P, and therefore its concentration in liquid In, depends on the substrate temperature [39].

**Optimization of the substrate temperature**

In order to better understand how the annealing temperature affects the MADSAG process, we have investigated the optical properties, morphology and crystalline quality of a set of nanoinsertions annealed at different temperatures under a PH$_3$ flux of 2.2x10$^{-2}$ mol/min. Fig. 3 compares the PL spectra acquired at 77K and 290K from the structures annealed at temperatures 550, 600 and 650°C. Details of the PL characterization are provided in section S5 of the SI. The PL signal from the three samples demonstrate similar shapes but have significantly different intensities. The highest PL intensity and the narrowest full width at half maximum (FWHM) are observed from the InP/Si structures annealed at 600°C, and the signals from the samples annealed at lower and higher temperatures both have decreased PL intensities. The signal from the sample annealed at 600 °C shows a narrow PL peak at the energy corresponding to the optical transition in the Γ point of InP, with FWHM below 25 meV at 77 K and 55 meV at 300 K, which is attributed to the high crystallinity of the grown InP structures.

Typical plan-view SEM images of arrays of the nanoinsertions after annealing at 550, 600 and 650°C are shown in Fig. 4a-c. We conceive that a square-base truncated pyramid shape of the structures in plan-view implies uniform growth of InP on all four etched {111} planes in Si and complete crystallization of In into InP. In the samples annealed at 600 °C (Fig. 4b) most of the structures have the truncated pyramidal shape. The structures annealed at slightly lower temperature (Fig. 4a) show similar morphology, while annealing at higher temperatures leads to irregularities in morphology (Fig. 4c). We relate the morphology variation to crystal defects and residual In after InP crystallization, which can exist under non-optimized growth conditions. In this case of non-optimal growth conditions, crystal defects tend to nucleate at the interface between the remaining In and InP. The crystallinity of these arrays of InP nanostructures were further examined by XRD.

Fig. 4d-f compares the XRD InP (111) pole figures acquired from the samples annealed at temperatures 550, 600 and 650°C, respectively. The peaks crystallographicaly aligned with the Si lattice are labeled with Miller indices, while the peaks arising from twinning in the InP crystal are marked with red arrows. The four-fold symmetry of the peaks from twinned structures is related to four possible symmetries inherited from nucleation on each of the four {111} Si planes formed in the mask openings. The samples annealed at different temperatures show very similar peak patterns.

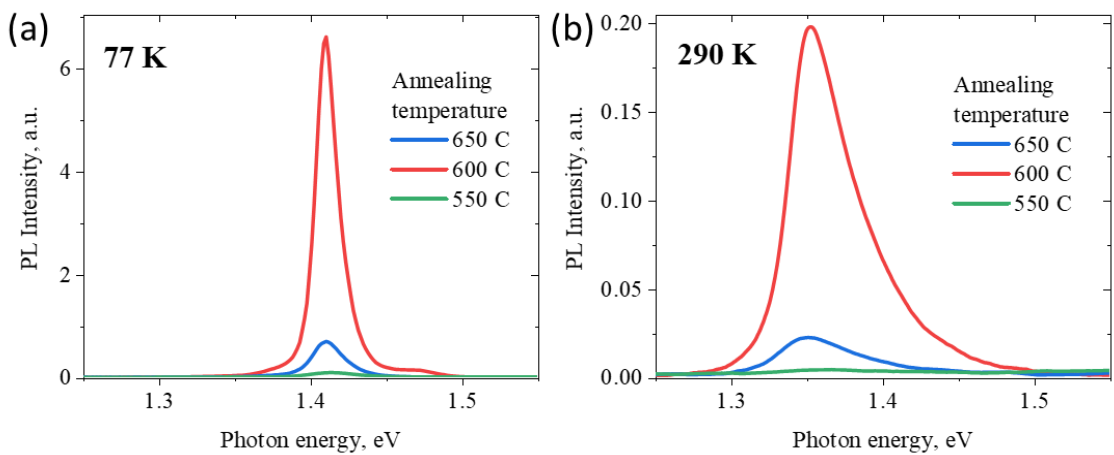

Fig. 3. Photoluminescence spectra acquired at 77 K (a) and 290 K (b) from samples annealed at a PH3 flux of 2.2 x 10-2 mol/min and at temperatures 550, 600 and 650 °C.

However, the comparison of the integrated intensity of the signal corresponding to the defect-free InP {111} planes, summarized in Table 1, shows that the samples annealed at 550 and 600 °C have higher intensity of these reflections and thus better crystalline qualities. The comparison of the data with a reference InP wafer can be found in section S6 of the SI.

XRD In (011) pole figures are displayed in Fig. 4g-i, revealing the impact of the annealing temperature on the presence of residual metallic In. The measurements from the three samples demonstrate similar peak positions corresponding to the four {110} crystallographic planes, while their intensities vary with annealing temperature (integrated reflection intensities are summarized in Table 1).

Table 1. Intensity of In (011) and InP (111) reflections

| Annealing temperature, °C | Integral intensity of In (011) reflection, a. u. | Integral intensity of InP (111) reflection, a. u. |
|---|---|---|
| 550 | 621 | 16910 |
| 600 | 250 | 17002 |
| 650 | 1243 | 6842 |

This shows that the residual In in the structures is highly crystalline with a bct crystal structure (space group I4/mmm), which is consistent with what we have observed in HR STEM images. Comparing the integral intensity of In (011) reflections in the pole figures from the three samples in Table 1, we conclude that the sample annealed at 650 °C has the largest amount of monocrystalline In, whereas the sample annealed at 600 °C has the least amount of residual In.

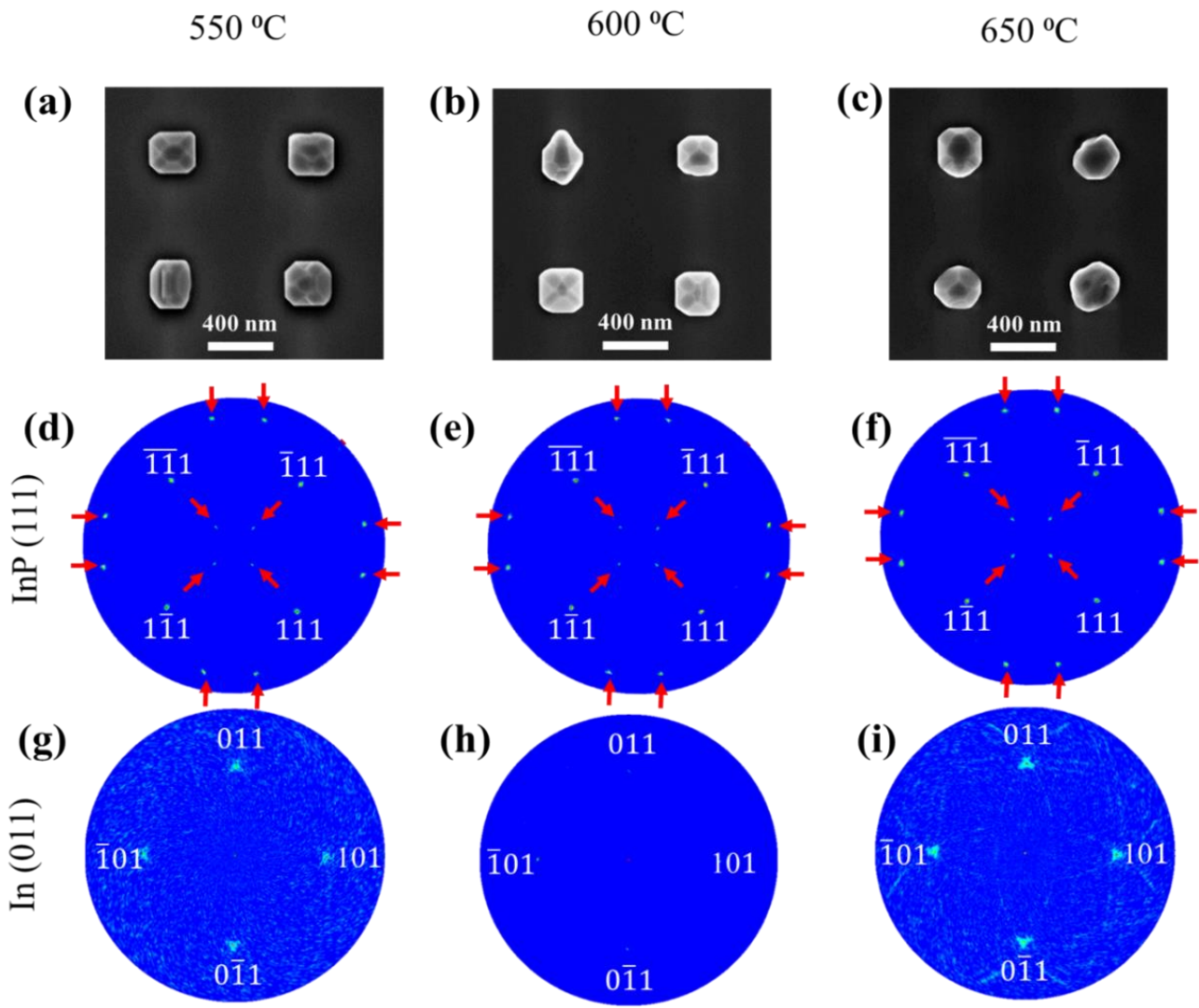

Fig. 4. Top-view SEM images of the InP structures annealed at 550 (a), 600 (b) and 650 °C (c). InP (111) (d-f) and In (011) (g-i) pole figures acquired from MADSAG InP/Si annealed at 550, 600 and 650 °C respectively.

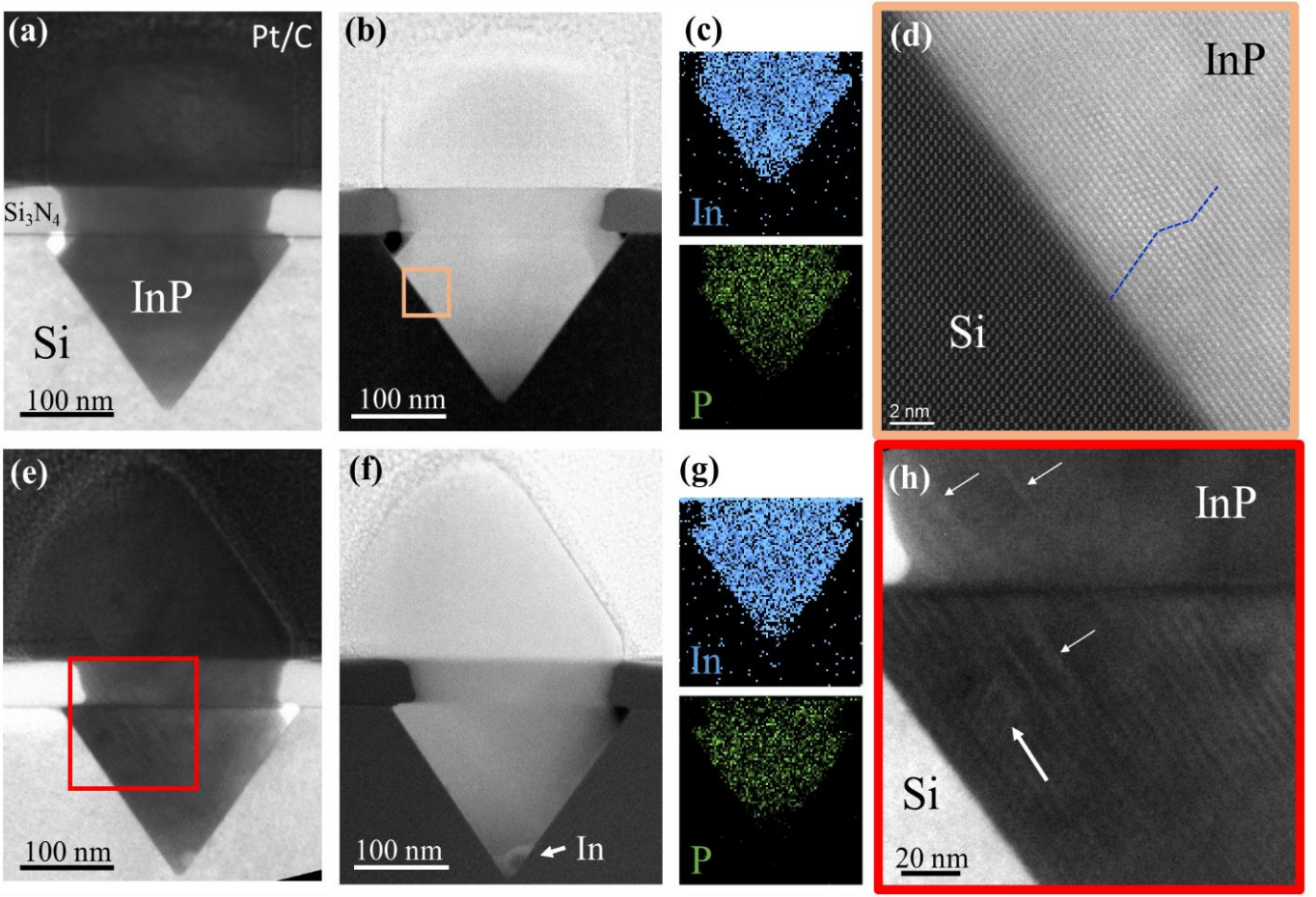

Fig. 5. Bright field TEM images (a, e and h), STEM images (b, f and d) and EDX maps (c and g) comparing InP nanoinsertions with symmetrical (a-d) and asymmetrical (e-h) plan-view profiles. (d) and (h) show zoomed views of the marked regions in (a) and (e), respectively. The dashed blue lines in (d) show a twinning in the InP crystal. The arrows in (h) indicate defects and dislocations in the InP crystal.

The results so far suggest that InP crystals having a symmetrical truncated octahedral pyramid profile in plan-view contain no residual In and are free from defects. Fig. 5 shows images of two structures in cross-sections, one with a such a profile (Fig. 5a-d) and the other with an irregular and asymmetrical plan-view profile (Fig. 5e-h). The EDX maps in Fig. 5c and 5g clearly confirm that the structure with an irregular plan-view shape contains residual In, while the structure with a truncated pyramid shape does not. Moreover, evidence of dislocations in the nanoinsertions with irregular plan-view profile can be seen in bright field TEM images in Fig. 5 e and h. Dislocations, however, are typically absent in the structures with truncated pyramid plan-view profiles and the only evidence of crystal twinning can be found in these structures, Fig. 5a and d.

We relate the incomplete consumption of In during InP formation with a lack of P atoms after full coverage of the top surface of the opening by InP crystal, which blocks further transport of P atoms to the In melt. This will be discussed further in the modeling describing the growth process.

We assume that the In(011) intensity in the XRD data, $I_{In(011)}$, is directly related to the residual number of In atoms $N_{In}^{res}$. We therefore introduce a growth model to estimate the amount of residual In and understand the non-monotonic temperature dependence of the In(011) XRD signal.

**Growth modeling**

The residual number of In atoms can be calculated as the difference between the initial number of In atoms in melt $N_{In}$ and the number of In atoms in solid InP. The number of In atoms in InP equals to the number of P atoms, which consist of P dissolved in liquid In before nucleation $N_P^0$ and absorbed during the crystallization $N_P^{cryst}$, thus:

$$N_{In}^{res} = N_{In} - N_P^0 - N_P^{cryst}. \qquad (1)$$

Nucleation and further crystallization of InP in the In melt are controlled by P concentration $C$ driven by the absorption and evaporation of P atoms. We assume that $C$ is rather uniform inside the In melt, due to the short characteristic time of the species mixing, which can be estimated as $\sqrt{A/D} \sim 10^{-3}$ s, where A is the area of the opening and $D$ is the diffusion coefficient of P in liquid In, which has the order of magnitude $\sim 10^{10} nm^2/s$ [40] at considered growth temperatures.

Before InP nucleation, P absorption at rate $F$ and evaporation at rate $C/\sigma\tau$ are balanced at the considered growth temperatures [39,41,42], and thus $C$ is given by the corresponding steady-state value $C_0 = \sigma F \tau$ and does not depend on time. In our notations, $\sigma$ is the area of the vapor-liquid interface per atom, and $\tau$ is the characteristic time of evaporation.

After nucleation, the amount of P adsorbed during crystallization can be determined by integrating the difference between the absorption and evaporation rates:

$$N_P^{cryst} = \int_0^{t_{cryst}} \left(F - \frac{C(t')}{\sigma\tau}\right) A dt'. \qquad (2)$$

Analytical calculations for $C(t)$ and $t_{cryst}$ given in the SI allow the determination of $N_P^0$ and $N_P^{cryst}$ as functions of the annealing temperature $T$. Using the rough estimation of the integral intensity of In (011) reflections $I_{In(011)} \sim N_{In}^{res}$, our modeling result can be summarized as:

$$I_{In(011)} = I_0 + I_1 exp\left[-\frac{T_{vap}}{T}\right] + I_2 exp\left[\frac{T_{vap}+T_D}{T}\right] \qquad (3)$$

With the temperature-independent term $I_0$, and increasing and decreasing exponential functions of the growth temperature with the corresponding prefactors $I_1$ and $I_2$. The temperature-dependent exponents are controlled by characteristic barriers for P evaporation $T_{vap}$ and diffusion $T_D$ in liquid In. Fig. 6 shows the integral intensity of the pole figure peaks corresponding to In(011) reflection fitted by Eq. (3) with fitting parameters $I_0$= - 127150, $I_1$=228333, $I_2$=85.7.

Eq. (3) brings together the effects of several processes, which contribute to the nonmonotonic temperature dependence of the residual amount of In, and thus the intensity of the In(011) XRD signal. The increasing exponent $I_1 exp[-T_{vap}/T]$ in Eq. (3) is proportional to $C_0$ and includes two contributions. First, according to our estimations (see section S7 in the SI), the P concentration before nucleation $C_0$ increases with temperature. Thus, less In remains at the end of the process, which corresponds to a decrease of the prefactor $I_1$ in Eq. (3) (see section S7 in the SI for details). Second, an increase in $C_0$ leads to faster motion of the growth front (shorter $t_{cryst}$). It may lead to the full coverage of the top surface of the opening by InP before the amount of P required for complete In conversion to InP gets absorbed. Thus, some residual In remains as we show in Fig. 5e, f. A similar situation can occur due to lowering the PH$_3$ flux during the annealing, as shown in Fig. 2c.

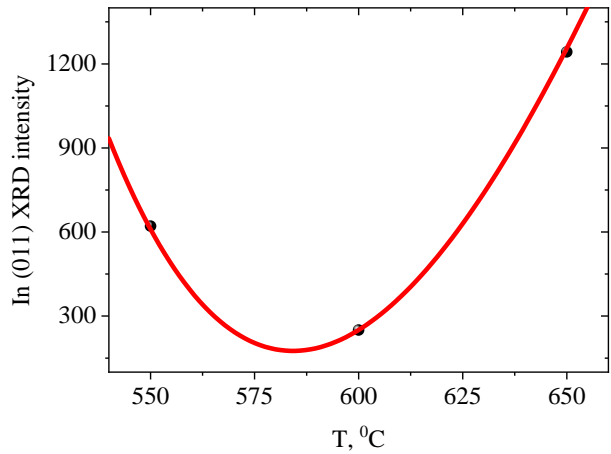

Fig. 6. The integral intensity of In(011) XRD signal $I_{In(011)}$ (dots) with the model fit described by Eq. (3) (solid line).

Meanwhile, during a faster crystallization stage less P evaporates from the melt and thus, less In remains, which is represented by the decreasing exponent $I_2 exp[(T_{vap}+T_D)/T]$ in Eq. (3). Therefore, the competing effects of shorter crystallization time and increasing initial P concentration shape the observed nonmonotonic temperature dependence.

**Conclusions**

In conclusion, we have proposed and investigated a new method called MADSAG for the selective-area epitaxial growth of InP on (100) Si. The InP nanostructures grown under optimized conditions show a very low crystal defect density and sharp interfaces between the grown InP and hosting Si. They also demonstrate a high intensity optical response, verified by PL measurements. We have investigated the influence of the growth parameters on the nucleation and morphology of InP nanoinsertions by SEM, HR STEM, EDX and XRD. Under non-optimal conditions, residual In remains in the structures, which negatively impacts the crystalline quality and the PL signal of the InP nanoinsertions. The remaining In can be eliminated and complete crystallization of the In melt can be achieved by optimizing the growth temperature and $PH_3$ flux. We thus conclude that MADSAG is a very promising CMOS compatible growth method, which can facilitate the development of novel active hybrid nanodevices integrated into the Si platform.

**References**

[1]  Teng M, Fathpour S, Safian R, Zhuang L, Honardoost A, Alahmadi Y, Polkoo S S, Kojima K, Wen H, Renshaw C K, LiKamWa P and Li G 2020 Miniaturized Silicon Photonics Devices for Integrated Optical Signal Processors *J. Light. Technol.* **38** 6–17

[2]  Tanaka Y 2020 High-speed and large-capacity integrated silicon photonics technologies *Metro and Data Center Optical Networks and Short-Reach Links III* ed M Glick, A K Srivastava and Y Akasaka (SPIE) p 19

[3]  Doerr C R 2015 Silicon photonic integration in telecommunications *Front. Phys.* **3** 1–16

[4]  del Alamo J A 2011 Nanometre-scale electronics with III–V compound semiconductors *Nature* **479** 317–23

[5]  Geum D-M, Park M-S, Lim J Y, Yang H-D, Song J D, Kim C Z, Yoon E, Kim S and Choi W J 2016 Ultra-high-throughput Production of III-V/Si Wafer for Electronic and Photonic Applications *Sci. Rep.* **6** 20610

[6]  Elshaari A W, Pernice W, Srinivasan K, Benson O and Zwiller V 2020 Hybrid integrated quantum photonic circuits *Nat. Photonics* **14** 285–98

[7]  Liao M, Chen S, Park J-S, Seeds A and Liu H 2018 III–V quantum-dot lasers monolithically grown on silicon *Semicond. Sci. Technol.* **33** 123002

[8]  Yu Y, Xue W, Semenova E, Yvind K and Mork J 2017 Demonstration of a self-pulsing photonic crystal Fano laser *Nat. Photonics* **11** 81–4

[9]  Aharonovich I, Englund D and Toth M 2016 Solid-state single-photon emitters *Nat. Photonics* **10** 631–41

[10]  Park G C, Xue W, Piels M, Zibar D, Mørk J, Semenova E and Chung I-S 2016 Ultrahigh-speed Si-integrated on-chip laser with tailored dynamic characteristics *Sci. Rep.* **6** 38801

[11]  Waldron N, Merckling C, Teugels L, Ong P, Ibrahim S A U, Sebaai F, Pourghaderi A, Barla K, Collaert N and Thean A V-Y 2014 InGaAs Gate-All-Around Nanowire Devices on 300mm Si Substrates *IEEE Electron Device Lett.* **35** 1097–9

[12]  Borg M, Schmid H, Moselund K E, Signorello G, Gignac L, Bruley J, Breslin C, Das Kanungo P, Werner P and Riel H 2014 Vertical III–V Nanowire Device Integration on Si(100) *Nano Lett.* **14** 1914–20

[13]  Kunert B, Mols Y, Baryshniskova M, Waldron N, Schulze A and Langer R 2018 How to control defect formation in monolithic III/V hetero-epitaxy on (100) Si? A critical review on current approaches *Semicond. Sci. Technol.* **33** 093002

[14]  Daix N, Uccelli E, Czornomaz L, Caimi D, Rossel C, Sousa M, Siegwart H, Marchiori C, Hartmann J M, Shiu K-T, Cheng C-W, Krishnan M, Lofaro M, Kobayashi M, Sadana D and Fompeyrine J 2014 Towards large size substrates for III-V co-integration made by direct wafer bonding on Si *APL Mater.* **2** 086104

[15]  Widiez J, Sollier S, Baron T, Martin M, Gaudin G, Mazen F, Madeira F, Favier S, Salaun A, Alcotte R, Beche E, Grampeix H, Veytizou C and Moulet J-S 2016 300 mm InGaAs-on-insulator substrates fabricated using direct wafer bonding and the Smart Cut™ technology *Jpn. J. Appl. Phys.* **55** 04EB10

[16]  Sahoo H K, Ottaviano L, Zheng Y, Hansen O and Yvind K 2018 Low temperature bonding of heterogeneous materials using $Al_2O_3$ as an intermediate layer *J. Vac. Sci. Technol. B, Nanotechnol. Microelectron. Mater. Process. Meas. Phenom.* **36** 011202

[17]  Sakanas A, Semenova E, Ottaviano L, Mørk J and Yvind K 2019 Comparison of processing-induced deformations of InP bonded to Si determined by e-beam metrology: Direct vs. adhesive bonding *Microelectron. Eng.* **214** 93–9

[18]  Osada A, Ota Y, Katsumi R, Kakuda M, Iwamoto S and Arakawa Y 2019 Strongly Coupled Single-Quantum-Dot–Cavity System Integrated on a CMOS-Processed Silicon Photonic Chip *Phys. Rev. Appl.* **11** 024071

[19]  Ludewig P, Reinhard S, Jandieri K, Wegele T, Beyer A, Tapfer L, Volz K and Stolz W 2016 MOVPE growth studies of Ga(NAsP)/(BGa)(AsP) multi quantum well heterostructures (MQWH) for the monolithic integration of laser structures on (001) Si-substrates *J. Cryst. Growth* **438** 63–9

[20]  Wan Y, Norman J C, Tong Y, Kennedy M J, He W, Selvidge J, Shang C, Dumont M, Malik A, Tsang H K, Gossard A C and Bowers J E 2020 1.3 μm Quantum Dot-Distributed Feedback Lasers Directly Grown on (001) Si *Laser Photon. Rev.* 2000037

[21]  Wan Y, Zhang S, Norman J C, Kennedy M J, He W, Liu S, Xiang C, Shang C, He J-J, Gossard A C and Bowers J E 2019 Tunable quantum dot lasers grown directly on silicon *Optica* **6** 1394

[22]	Luxmoore I J, Toro R, Pozo-Zamudio O Del, Wasley N A, Chekhovich E A, Sanchez A M, Beanland R, Fox A M, Skolnick M S, Liu H Y and Tartakovskii A I 2013 III-V quantum light source and cavity-QED on silicon *Sci. Rep.* **3** 1–5
[23]	Mauthe S, Vico Trivino N, Baumgartner Y, Sousa M, Caimi D, Stoferle T, Schmid H and Moselund K E 2019 InP-on-Si Optically Pumped Microdisk Lasers via Monolithic Growth and Wafer Bonding *IEEE J. Sel. Top. Quantum Electron.* **25** 1–7
[24]	Merckling C, Waldron N, Jiang S, Guo W, Barla K, Heyns M, Collaert N, Thean A and Vandervorst W 2014 Selective-Area Metal Organic Vapor-Phase Epitaxy of III-V on Si: What About Defect Density? *ECS Trans.* **64** 513–21
[25]	Wang Z, Tian B, Pantouvaki M, Guo W, Absil P, Van Campenhout J, Merckling C and Van Thourhout D 2015 Room-temperature InP distributed feedback laser array directly grown on silicon *Nat. Photonics* **9** 837–42
[26]	Vukajlovic-Plestina J, Kim W, Ghisalberti L, Varnavides G, Tütüncuoglu G, Potts H, Friedl M, Güniat L, Carter W C, Dubrovskii V G and Fontcuberta i Morral A 2019 Fundamental aspects to localize self-catalyzed III-V nanowires on silicon *Nat. Commun.* **10** 869
[27]	Bollani M, Bietti S, Frigeri C, Chrastina D, Reyes K, Smereka P, Millunchick J M, Vanacore G M, Burghammer M, Tagliaferri A and Sanguinetti S 2014 Ordered arrays of embedded Ga nanoparticles on patterned silicon substrates *Nanotechnology* **25** 205301
[28]	Oehler F, Cattoni A, Scaccabarozzi A, Patriarche G, Glas F and Harmand J-C 2018 Measuring and Modeling the Growth Dynamics of Self-Catalyzed GaP Nanowire Arrays *Nano Lett.* **18** 701–8
[29]	Mayer B F, Wirths S, Mauthe S, Staudinger P, Sousa M, Winiger J, Schmid H and Moselund K E 2019 Microcavity Lasers on Silicon by Template-Assisted Selective Epitaxy of Microsubstrates *IEEE Photonics Technol. Lett.* **31** 1021–4
[30]	Wang Z, Tian B, Pantouvaki M, Guo W, Absil P, Van Campenhout J, Merckling C and Van Thourhout D 2015 Room-temperature InP distributed feedback laser array directly grown on silicon *Nat. Photonics* **9** 837–42
[31]	Herranz J, Corfdir P, Luna E, Jahn U, Lewis R B, Schrottke L, Lähnemann J, Tahraoui A, Trampert A, Brandt O and Geelhaar L 2020 Coaxial GaAs/(In,Ga)As Dot-in-a-Well Nanowire Heterostructures for Electrically Driven Infrared Light Generation on Si in the Telecommunication O Band *ACS Appl. Nano Mater.* **3** 165–74
[32]	Kunert B, Guo W, Mols Y, Tian B, Wang Z, Shi Y, Van Thourhout D, Pantouvaki M, Van Campenhout J, Langer R and Barla K 2016 III/V nano ridge structures for optical applications on patterned 300 mm silicon substrate *Appl. Phys. Lett.* **109** 091101
[33]	Kunert B, Mols Y, Baryshniskova M, Waldron N, Schulze A and Langer R 2018 How to control defect formation in monolithic III/V hetero-epitaxy on (100) Si? A critical review on current approaches *Semicond. Sci. Technol.* **33** 093002
[34]	Güniat L, Caroff P and Fontcuberta i Morral A 2019 Vapor Phase Growth of Semiconductor Nanowires: Key Developments and Open Questions *Chem. Rev.* **119** 8958–71
[35]	Tauchnitz T, Berdnikov Y, Dubrovskii V G, Schneider H, Helm M and Dimakis E 2018 A simple route to synchronized nucleation of self-catalyzed GaAs nanowires on silicon for sub-Poissonian length distributions *Nanotechnology* **29** 504004
[36]	Wang J, Sciarrino F, Laing A and Thompson M G 2020 Integrated photonic quantum technologies *Nat. Photonics* **14** 273–84
[37]	Reyes K, Smereka P, Nothern D, Millunchick J M, Bietti S, Somaschini C, Sanguinetti S and Frigeri C 2013 Unified model of droplet epitaxy for compound semiconductor nanostructures: Experiments and theory *Phys. Rev. B* **87** 165406
[38]	Sanguinetti S, Bietti S and Koguchi N 2018 Droplet Epitaxy of Nanostructures *Molecular Beam Epitaxy* (Elsevier) pp 293–314
[39]	Bachmann K J and Buehler E 1974 Phase Equilibria and Vapor Pressures of Pure Phosphorus and of the Indium/Phosphorus System and Their Implications Regarding Crystal Growth of InP *J. Electrochem. Soc.* **121** 835
[40]	Srnánek R and Hábovčík P 1979 A method for diffusion coefficient determination of phosphorus in liquid indium *J. Cryst. Growth* **46** 55–8
[41]	Ansara I, Chatillon C, Lukas H L, Nishizawa T, Ohtani H, Ishida K, Hillert M, Sundman B, Argent B B, Watson A, Chart T G and Anderson T 1994 A binary database for III–V compound semiconductor systems *Calphad* **18** 177–222
[42]	Kuphal E 1984 Phase diagrams of InGaAsP, InGaAs and InP lattice-matched to (100)InP *J. Cryst. Growth* **67** 441–57